\documentclass[12pt]{article}
\usepackage{amsmath}
\def\be{\begin{equation}}
\def\ee{\end{equation}}
\def\arr{\begin{array}{rll}}
\def\ea{\end{array}}
\def\bea{\begin{eqnarray}}
\def\eea{\end{eqnarray}}

\def\N2{$N{=}2$}

\def\>{\rangle}
\def\<{\langle}
\def\+{\dagger}
\def\={\ =\ }

\begin{document}
\renewcommand{\thefootnote}{\fnsymbol{footnote}}
\begin{titlepage}
\setcounter{page}{0}
\vskip 1cm
\begin{center}
{\LARGE\bf  Comment on "Generalized near horizon  }\\
\vskip 0.5cm
{\LARGE\bf    extreme binary black hole geometry" }\\
\vskip 2cm
$
\textrm{\Large Anton Galajinsky \ }^{a,b}
$
\vskip 0.7cm
${}^{a}$ {\it
Tomsk Polytechnic University,
634050 Tomsk, Lenin Ave. 30, Russia} \\
\vskip 0.2cm
{\it
${}^{b}$ Tomsk State University of Control Systems and Radioelectronics, 634050 Tomsk, Lenin Ave. 40, Russia} \\
\vskip 0.2cm
{e-mail: galajin@tpu.ru}

\end{center}
\vskip 1cm
\begin{abstract} \noindent
It is demonstrated that the near--horizon geometry of two extreme Kerr black holes of equal mass, which are held a finite distance apart by a massless strut, introduced recently in [Phys. Rev. D 100 (2019) 044033], is a particular member of the near horizon Kerr--Bolt class.
\end{abstract}

\vspace{0.5cm}

\end{titlepage}

\renewcommand{\thefootnote}{\arabic{footnote}}
\setcounter{footnote}0

The near horizon geometries of the extreme black holes in diverse dimensions attracted recently considerable attention (for a review see \cite{KL}). In this context, the key feature is the enhanced isometry group which involves the conformal factor $SL(2,R)$. Denoting the temporal, radial, and azimuthal coordinates by $t$, $r$, and $\phi_i$, $i=1,\dots,n$, one can represent the $SL(2,R)$--transformations in the form
\begin{align}\label{1}
&
t'=t+\alpha; &&  &&
\nonumber\\[2pt]
&
t'=t+\beta t, && r'=r-\beta r; &&
\nonumber\\[2pt]
&
t'=t+(t^2+\frac{1}{r^2}) \gamma, &&r'=r-2 tr\gamma, &&
\phi'_i=\phi_i-\frac{2}{r} \gamma,
\end{align}
where the infinitesimal parameters $\alpha$, $\beta$, and $\gamma$ correspond to the time translation, dilatation and special conformal transformation, respectively. The latitudinal coordinates remain inert under the action of the conformal group. The transformations (\ref{1}) proved crucial for various physical applications including the Kerr/CFT--correspondence \cite{GHSS}, the construction of novel superconformal mechanics models \cite{G2,G1}, and the study of new integrable systems associated with black hole backgrounds \cite{HNS,DNSS}.

Typically, a near horizon geometry is constructed by implementing a specific limit to
a given extreme black hole configuration \cite{KL}. Yet, one can turn the logic around and use the
conformal invariants associated with (\ref{1})
\be\label{ci}
r^2 dt^2-\frac{dr^2}{r^2}, \qquad r dt+d \phi_i, \qquad d \phi_i-d\phi_j,
\ee
so as to build a Ricci--flat metric. It suffices to consider the most general quadratic form built in terms of (\ref{ci}) and involving arbitrary coefficient functions depending on the latitudinal coordinates only, and to impose the Einstein equations. In four dimensions, the analysis can be carried out in full generality \cite{GO} which reproduces the near horizon Kerr--Bolt metric \cite{Gh}
\bea\label{metr}
&&
ds^2=a(\theta)\left( -r^2 dt^2+\frac{dr^2}{r^2}+d\theta^2\right)+b(\theta) {\left(r dt+d \phi \right)}^2,
\nonumber\\[2pt]
&&
a(\theta)=L_1 (1+\cos^2{\theta})+L_2 \cos{\theta}, \qquad b(\theta)=\frac{(4 L_1^2 - L_2^2) \sin^2{\theta}}{a(\theta)},
\eea
where $\theta$ is the latitudinal angular variable and $L_1$, $L_2$ are free parameters.\footnote{In order to guarantee the Lorentzian signature, one has to demand $4 L_1^2>L_2^2$.}
Thus, the $SL(2,R)$ invariance alone allows one to unambiguously fix the NUT--charge extension of the $4d$ near horizon extreme Kerr (NHEK) geometry \cite{BH}.

In a very recent work \cite{CHRR} (see also \cite{CHRR1}), the near--horizon geometry of two extreme Kerr black holes of equal mass, which are held a finite distance apart by a massless strut, was introduced
\bea\label{NHB}
&&
ds^2=\Gamma(\Theta)\left[-R^2 dT^2+\frac{dR^2}{R^2}+d\Theta^2+\Lambda^2(\Theta){\left(d\Phi+\frac{\sqrt{11}-\sqrt{3}}{2} R dT\right)}^2 \right],
\nonumber\\[2pt]
&&
\qquad \quad
\Gamma(\Theta)=\frac{2(3\sqrt{33}-13)\cos{\Theta}+(15-\sqrt{33})(3+\cos{2\Theta})}{16},
\nonumber\\[2pt]
&&
\qquad \quad
\Gamma(\Theta) \Lambda^2(\Theta)=\frac{256 \sin^2{\Theta}}{4(-59+11\sqrt{33})\cos{\Theta}+(93-13\sqrt{33})(3+\cos{2\Theta})}.
\eea
Because the parent metric \cite{MR} was rather complicated, the authors of \cite{CHRR} fixed free parameters of the original binary system so as to implement the near horizon limit in the most tractable way.
The line element (\ref{NHB}) was interpreted as describing the NHEK black hole pierced by the cosmic string/conical singularity, which extends all
the way from the horizon to infinity. It was called the pierced--NHEK geometry \cite{CHRR}.

It was also argued in \cite{CHRR} that (\ref{NHB}) has the $SL(2,R)\times U(1)$ isometry group. Yet, if (\ref{metr}) were the most general metric constructed from the $SL(2,R)$ invariants, (\ref{NHB}) would belong to that class. Let us demonstrate that this is indeed the case.

Taking into account the elementary relations $3+\cos{2\Theta}=2(1+\cos^2{\Theta})$, $\frac{93 - 13 \sqrt{33}}{15-\sqrt{33}} -
\frac{ 2 (-59 + 11 \sqrt{33})}{3 \sqrt{33} - 13}=0$, redefining the azimuthal angular variable $\frac{2 \Phi}{\sqrt{11}-\sqrt{3}}=\tilde\Phi$, and identifying the coordinates $(T,R,\Theta,\tilde\Phi)$ with $(t,r,\theta,\phi)$ above, one finds that (\ref{NHB}) is a particular member of the two--parameter family (\ref{metr}) which occurs at
\be
L_1=\frac{15-\sqrt{33}}{8}, \qquad L_2=\frac{3\sqrt{33}-13}{8}.
\ee

Thus, we have demonstrated that the pierced--NHEK geometry introduced recently in \cite{CHRR} is a particular member of the well known near horizon extreme Kerr--Bolt class \cite{Gh}.

\vspace{0.5cm}

\noindent{\bf Acknowledgements}\\

\noindent
This work was supported by the RFBR grant 18-52-05002 and the Tomsk Polytechnic University competitiveness enhancement program.

\end{document}